\documentclass{emulateapj}
\usepackage{graphicx}
\usepackage{amssymb,amsmath}
\usepackage{color}

\shortauthors{Coppola et al.}

\begin{document}


\title{Non-equilibrium H$_2$ formation in the early Universe:\\
energy exchanges, rate coefficients and spectral distortions}
\shorttitle{Non-equilibrium H$_2$ formation in the early Universe}


\author{C. M. Coppola\altaffilmark{1,2}, R. D'Introno\altaffilmark{3}, 
D. Galli\altaffilmark{4}, J. Tennyson\altaffilmark{2}, S. Longo\altaffilmark{1,5}}
\altaffiltext{1}{Universit\`a degli Studi di Bari, Dipartimento di Chimica, Via Orabona 4, I-70126 Bari, Italy}
\altaffiltext{2}{Department of Physics and Astronomy, University College London, Gower Street, London WC1E 6BT}
\altaffiltext{3}{Universit\`a degli Studi di Bari, Dipartimento di Fisica, Via Amendola 173, I-70126 Bari, Italy}
\altaffiltext{4}{INAF-Osservatorio Astrofisico di Arcetri, Largo E.~Fermi 5, I-50125 Firenze, Italy}
\altaffiltext{5}{IMIP-CNR, Section of Bari, via Amendola 122/D, I-70126 Bari, Italy}
\email{carla.coppola@chimica.uniba.it}


\begin{abstract}
Energy exchange processes play a crucial role in the early Universe,
affecting the thermal balance and the dynamical evolution of the
primordial gas. In the present work we focus on the consequences of a
non-thermal distribution of the level populations of H$_2$: first, we
determine the excitation temperatures of vibrational transitions and the
non-equilibrium heat transfer; second, we compare the modifications to
chemical reaction rate coefficients with respect to the values obtained
assuming local thermodynamic equilibrium; third, we compute the
spectral distortions to the cosmic background radiation generated by
the formation of H$_2$ in vibrationally excited levels. We conclude
that non-equilibrium processes cannot be ignored in cosmological
simulations of the evolution of baryons, although their observational
signatures remain below current limits of detection. New fits to the equilibrium and non-equilibrium heat transfer functions are provided.
\end{abstract}


\section{Introduction}

Understanding the thermal evolution of the Universe in the epoch in
which atoms and molecules formed is a crucial step to properly model
the birth of the first bound structures (e.g., Flower \& Pineau des For{\^e}ts~2001). In particular, the balance
between cooling and heating processes has to be taken into account and
modeled according to the chemical and physical processes occuring in
the primordial plasma. It is well established that Lyman-$\alpha$
cooling is effective at gas temperature higher than $\sim 8000$~K,
corresponding to redshifts $z\gtrsim 2700$,  while primordial
molecules, in particular H$_2$ and HD formed at $z\lesssim 1000$, are
the most efficient cooling agents of the pristine plasma at lower
temperatures.

Several authors have calculated the heating and cooling functions of the primordial
molecular species: \cite{b12}, Lepp \& Shull~(1984), \cite{b44}, \cite{b4}, \cite{b11}, Galli \&
Palla~1998 (hereafter GP98), \cite{b15}.  One of the standard
assumptions in these calculations is that the population of internal
states can be described by a Boltzmann distribution. This hypothesis is
valid in many astrophysical environments where the density is
sufficiently high to bring the internal degrees of freedom to a
condition of local thermodynamic equilibrium (LTE). For a given
species, e.g. H$_2$, this condition is quantified in terms of a
critical density $n_{\rm cr}({\rm H})$ defined as the ratio between
radiative and collisional de-excitation coefficients of H$_2$. It is
easy to check that even at highest redshifts  ($z \approx 1000$),
the ambient baryon density $n_{\rm b}$ is $\sim 2$ orders of magnitude
below the critical density $n_{\rm cr}({\rm H})$ (although considering specific rotational transitions the critical density is below the baryon one up to lower redshift $z \approx 500$).  In addition, several
physical phenomena (e.g. shocks) and chemical processes can produce
deviations from LTE.  In particular, most of the gas-phase molecular
formation processes selectively produce species in states that deviate
significantly from LTE.

In the case of the early Universe, \cite{b2} (hereafter C11) computed the vibrational
distribution of H$_2$ and H$_2^+$ formed at redshifts $10<z<1000$ and
found that high sovrathermal tails are present especially at low $z$. The
existence of non-equilibrium features in the level populations of H$_2$
is important, because of the role of this species as a coolant of
primordial gas. In the present work, we extend the work of C11 to
study the following physical quantities relevant to the non-equilibrium energy
exchange in the primordial Universe: excitation temperatures, heat transfer
functions, reaction rates and spectral distortions of cosmic background
radiation (CBR).

\section{Reduced model: ortho- and para- states}

Because the number of roto-vibrational levels involved in the kinetics
of H$_2$ is too high ($\approx 300$) for a direct extension of the approach
used by C11, in the present work we implement a reduced
model in order to provide a simpler starting point for more extended
calculations. This model is based on the assumptions that: (1) the most
important channels determining the population of vibrational states are
the associative detachment reaction 
\[
{\rm H}^- + {\rm H}\rightarrow {\rm H}_2+e,
\]
and the spontaneous and stimulated radiative transitions between 
roto-vibrational levels; and
(2) a steady-state approximation can be applied to the kinetics of
vibrational levels.  Under these hypotheses, two steady-state Master
Equations (for ortho- and para- states, respectively) can be written
for the 15 vibrational levels ($i=0,14$) of H$_2$:
\begin{equation}
f_i \sum_{j,i\neq j} R_{ij} - \sum_{j,i\neq j} R_{ji} f_j = 
k_i f_{\textrm{H}^-}f_{\textrm{H}}n_b,
\label{masterequation}
\end{equation}
where $f_i$ is the fractional abundance of H$_2$ in the $i^{\rm th}$
level, $f_{\textrm{H}^-}$ and $f_\textrm{H}$ those of H$^-$ and H
respectively, $R_{ij}$ is the matrix of radiative coefficients
including absorption processes, calculated as in C11 averaging
over the initial rotational levels and summing over the final ones the
Einstein coefficients computed by \cite{b9}. The values of
$f_{\textrm{H}^-}$ and $f_{\textrm{H}}$ are taken from the complete
kinetic model by C11. The rate coefficients $k_i$ for the
associative detachment reaction for the $i^{\rm th}$ vibrational level
formation have been evaluated using the
cross-sections $\sigma_{ij}$ calculated by \cite{b19} summing over all
final rotational states:
\begin{equation}
k_i(T_{\rm m})=\sum_{j=0}^{j_{\rm max}(i)} \tilde k_{ij}(T_{\rm m}),
\label{ki}
\end{equation}
where
\begin{equation}
\tilde k_{ij}(T_{\rm m})=\sqrt{\frac{8}{\pi\mu(k_{\rm B}T_{\rm m})^3}}
\int_0^\infty E\, \sigma_{ij}(E) e^{-E/(k_{\rm B}T_{\rm m})}\, dE,
\label{kij}
\end{equation}
with $\mu$ reduced mass of the system and $T_{\rm m}$ matter temperature. It should be noted that
experimental results by \cite{b8} recently showed very good
agreement with these quantum calculations.  

\begin{figure}
\includegraphics[width=6cm,angle=-90]{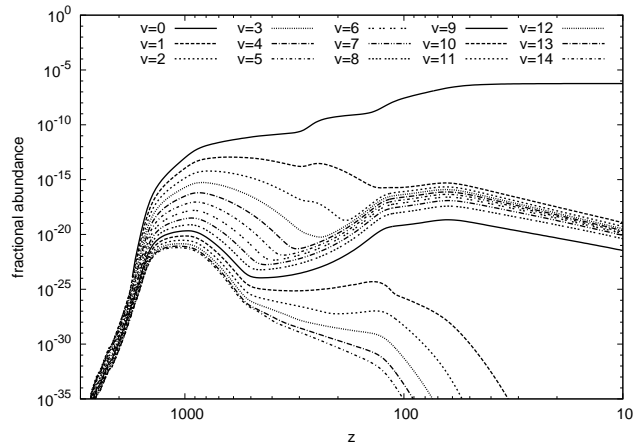} 
\caption{Fractional abundances of the vibrational levels of H$_2$ 
according to the reduced steady-state model
described in the text.}
\label{f1}
\end{figure}

The equation for the vibrational ground state $f_0$ in
Eq.~\ref{masterequation} is replaced by the normalization condition
\begin{equation}
\sum_i f_i = f_{\textrm{H}_2},
\end{equation}
where $f_{\textrm{H}_2}(z)$ is the fraction of H$_2$ which is also
taken from the complete model. Figure~\ref{f1} shows the fractional
abundance of vibrational levels obtained using the reduced model
described in the present section. The results for all $f_i$ are
satisfactory close to those obtained with the fully kinetic model shown
in Figure~10 of C11, at least for not too high values of $z$.
These results show that the hypothesis of steady-state can be applied
to the present problem with enough confidence to proceed to the study
of the ortho- and para- states.  For this  we calculate two sets of
rate coefficients $k_{i,\textrm{ortho}}(T_{\rm m})$ and
$k_{i,\textrm{para}}(T_{\rm m})$ for each vibrational level $i$,
\begin{equation}
\begin{split}
&k_{i,\textrm{ortho}}(T_{\rm m})=\sum_{j~\textrm{odd}} \tilde k_{ij}(T_{\rm m}),\\
&k_{i,\textrm{para}}(T_{\rm m})=\sum_{j~\textrm{even}} \tilde k_{ij}(T_{\rm m}).
\end{split}
\end{equation}
The $R_{ij}$ coefficients are also thermally averaged over a partial distribution. 
The first equation of each system is replaced by the normalization condition
\begin{equation}
\sum_i f_{i,\textrm{ortho/para}} = 1,
\end{equation}
which means that we are calculating the vibrational distribution of
each of the two species but not the total fraction of ortho- and para-
hydrogen, which cannot be calculated by a steady-state approach.

\section{Excitation temperatures}

For each transition $0$--$v$, the excitation temperature is defined as
\begin{equation}
T_{0-v}=\frac{E_v-E_0}{k_{\rm B}\ln(n_0/n_v)},
\end{equation}
where $E_v$ and $n_v$ represent the energy and fractional abundance of
the $v^{\rm th}$ vibrational level, respectively, and $k_{\rm B}$ is
the Boltzmann constant. 

\begin{figure}
\begin{center}$
\begin{array}{l}
\includegraphics[width=9cm]{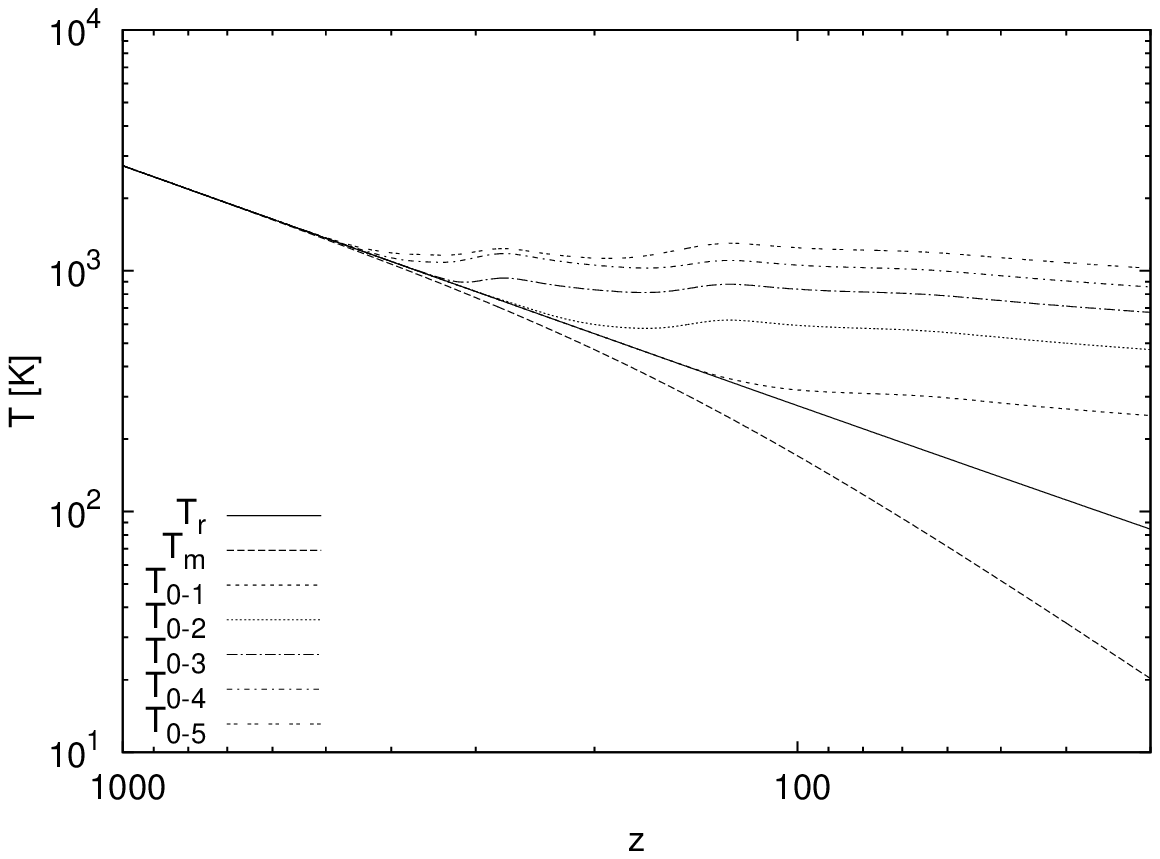} \\
\includegraphics[width=9cm]{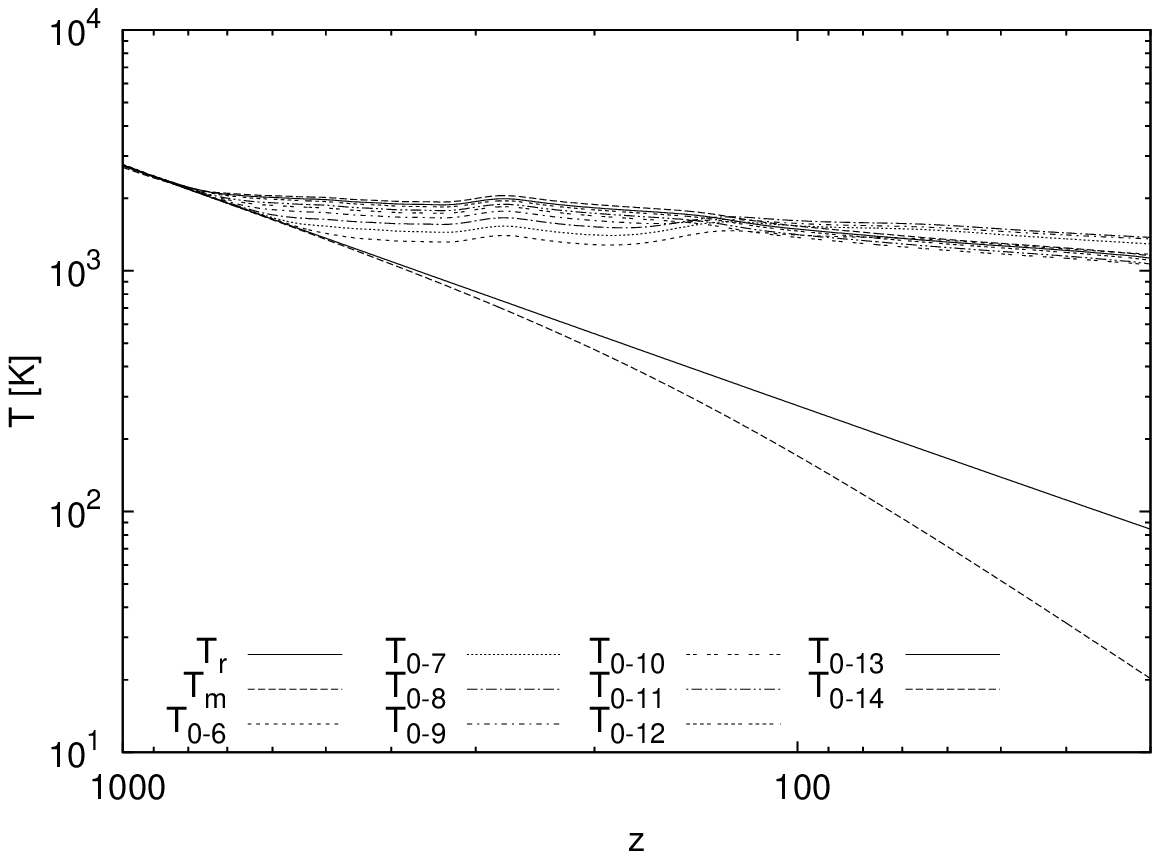} 
\end{array}$
\end{center}
\caption{H$_2$ excitation temperatures $T_{0-v}$, compared to the
temperature of the radiation ({\it solid curve}\/) and matter ({\it
dashed curve}\/).  {\it Top panel}: $v=1$ to $v=5$; {\it bottom panel}:
$v=6$ to $v=14$.}
\label{f2}
\end{figure}

In Figure~\ref{f2}, the excitation temperature of the transitions $0-v$
is compared to the temperatures of matter and radiation as a function
of $z$. For better clarity, a few curves have been repeated in the two
panels. The figure shows that vibration decouples from the
translational motion, and this happens at higher $z$ than the decoupling
of radiation and matter; for the most excited levels this corresponds
to an epoch where the conditions are near equilibrium. Later, after the
decoupling of matter and radiation, $T_{0-1}$ is slightly larger than the radiation temperature $T_{\rm r}$;
 the excitation temperatures $T_{0-v}$ for higher $v$ are
progressively higher. This result can be explained by considering the
reaction network: the vibrational levels are formed by associative
reactions that preferentially populate highly excited levels, while the
vibrational manifold as a whole is coupled to the CBR (which provides a
heat sink) much better than to the matter; this latter coupling occurs
via  the relatively ineffective H$_2$/H VT processes. Therefore, all
levels pairs are expected to be warmer than the radiation field, but
the lowest pairs are closer to equilibrium with the radiation because the chemical heating is lower.


The separation of $T_{0-1}$ and $T_{\rm r}$ occurs at $z\approx 100$,
which brings this phenomenon not far from potential indirect
observation (e.g. effects on reaction rates of processes and consequent different fractional abundances of chemical species at lower $z$). Another relevant feature of our calculation is that
$T_{0-1}$ is stable at about 200~K at the age of formation of the first
structures. Since the $1\rightarrow 0$ transition is an important heat
radiator, our results indicate that $T_{0-1}$ is a more appropriate
initial condition for the vibrational temperature of H$_2$ in
hydrodynamic collapse models than $T_{\rm r}$, which is considerably
lower. In Table~\ref{tab1} the values for $z_{\rm dec}$ (redshift at which excitation temperature decouples from radiation temperature) and $z_{\rm freeze-out}$ (redshift at which the freeze-out temperature is reached) are reported for each $T_{0-i}$; the former are evaluated considering a
relative deviation from $T_{\rm r}$ larger than 1\%, while the latter
are calculated searching for relative deviation of $T_{0-i}$ at each
$z$ from the value at $z=10$ smaller than $10^{-3}$.

\begin{table}
\begin{center}
\caption{Decoupling redshift of excitation temperatures}
\begin{tabular}{lll}
\hline
Excitation temperature (K)& $z_{\rm dec}$& $z_{\rm freeze-out}$\\ 
\hline
$T_{0-14}$ & 697 & 382\\
$T_{0-13}$ & 689 & 379\\
$T_{0-12}$ & 676 & 377\\
$T_{0-11}$ & 647 & 379\\
$T_{0-10}$ & 627 & 382\\
$T_{0-9}$  & 593 & 537\\
$T_{0-8}$  & 551 & 524\\
$T_{0-7}$  & 511 & 389\\
$T_{0-6}$  & 457 & 389\\
$T_{0-5}$  & 391 & 377\\
$T_{0-4}$  & 363 & 346\\
$T_{0-3}$  & 300 & 313\\
$T_{0-2}$  & 198 & 174\\
$T_{0-1}$  & 108 & 84\\
\hline
\label{tab1}
\end{tabular}
\end{center}
\end{table}

\begin{figure}
\begin{center}$
\begin{array}{l}
\includegraphics[width=6.4cm,angle=-90]{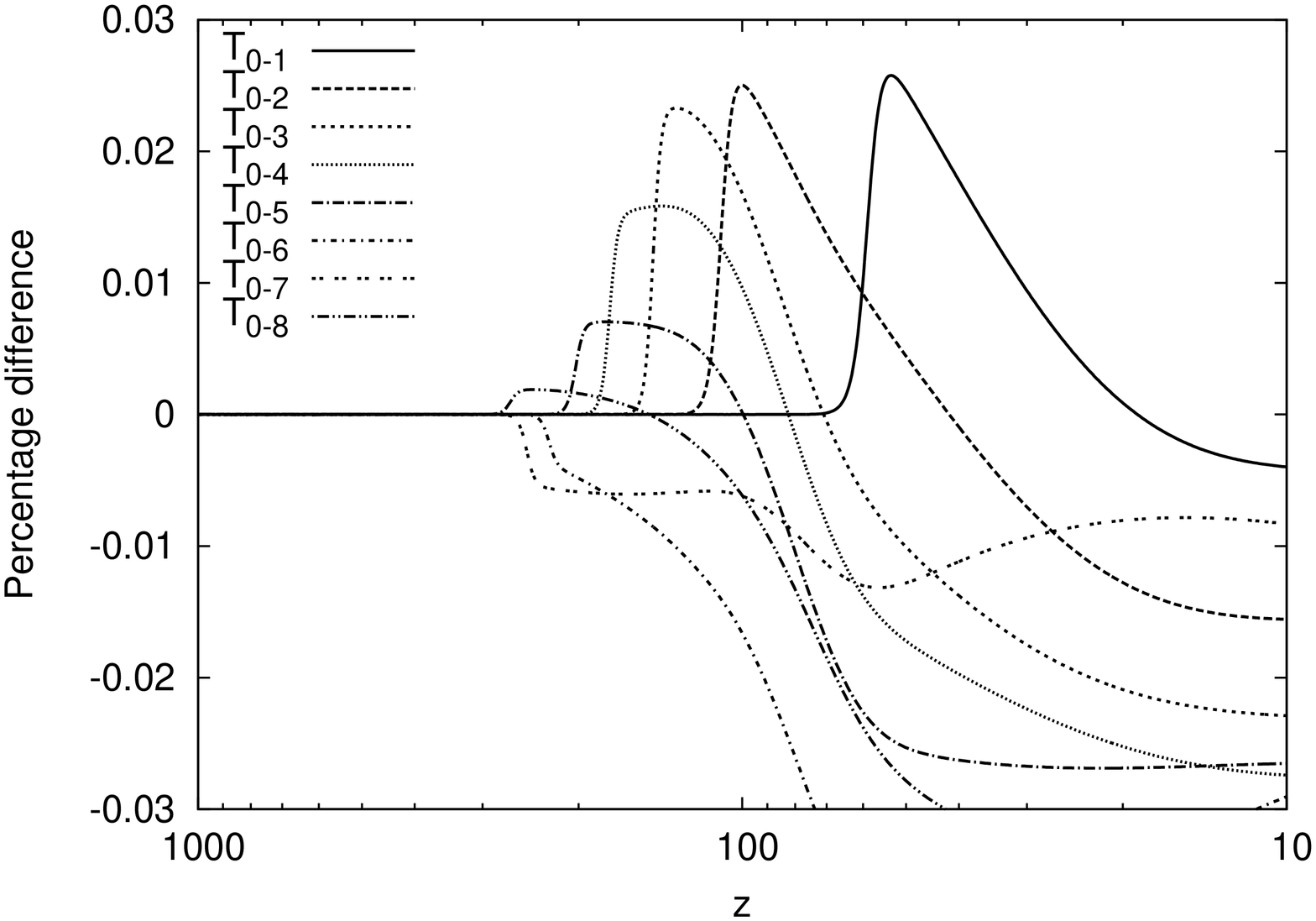}\\ 
\includegraphics[width=6.4cm,angle=-90]{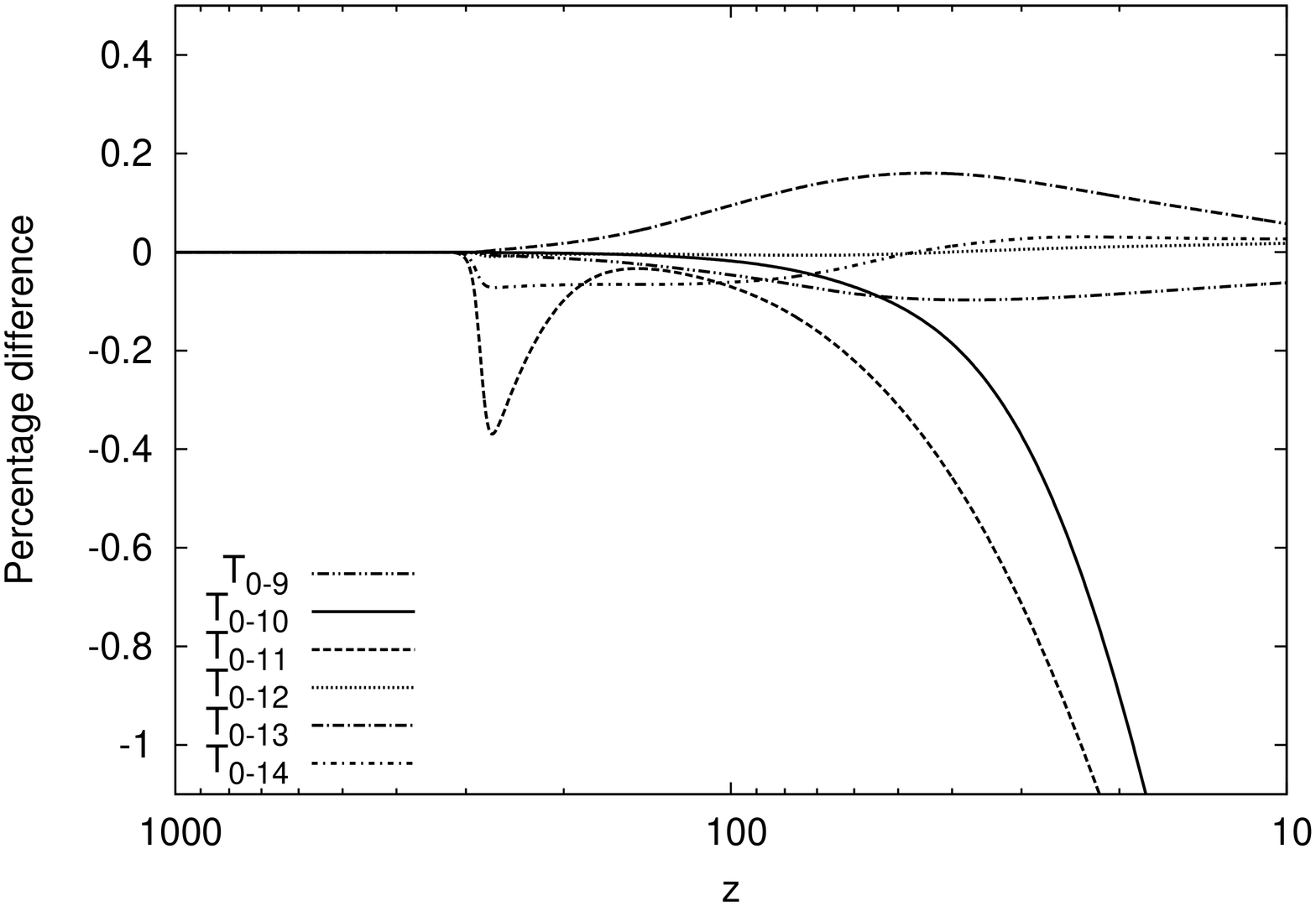} 
\end{array}$
\end{center}
\caption{Ratio of excitation temperatures for ortho- and para- states
given as ($T_{0-i,\textrm{ortho}}-T_{0-i,\textrm{para}})/T_{0-i,\textrm{ortho}}$.
{\it Top panel}: $v=1$ to $v=8$; {\it bottom panel}:
$v=9$ to $v=14$.}
\label{f3}
\end{figure}

Figure~\ref{f3} shows the values of the relative difference between
the vibrational excitation temperatures of ortho- and para- states. As
can be seen, significant deviations between vibrational temperature of
states with different rotational symmetry can be detected at low $z$
and high $i$. These differences suggest that the issue of rotational
non-equilibrium could be important and deserves to be addressed
accurately in future studies, i.e. by solving the Master Equation for a
full rotovibrational manifold. 


\section{Heat transfer function}
In this paper we discuss the role of chemical energy from exothermic reaction, which is ultimately dissipated either into radiation or into the thermal energy of H atoms. This energy flow is described by heating and cooling functions, usually indicated with the symbols $\Gamma$ and $\Lambda$, respectively. We consider the net molecular heat transfer, defined as the sum of all radiative excitations of H$_2$ followed by collisional de-excitations with H atoms and all collisional excitations followed by radiative decay:
\begin{equation}\begin{split}
&\Phi(T_{\rm m},T_{\rm r})=(\Gamma-\Lambda)_{\rm H_2}=\frac{1}{n({\rm H}_2)} \times \\
& \sum_{(v',j')<(v,j)}(n_{(v',j')}(T_{\rm r})\cdot k_{(v',j')\rightarrow(v,j)}(T_{\rm m})\\
&-n_{(v,j)}(T_{\rm r})\cdot k_{(v,j)\rightarrow(v',j')}(T_{\rm m})
(E_{v,j}-E_{v',j'}),
\label{cooling}
\end{split}\end{equation}
where $T_{\rm m}$ and $T_{\rm r}$ are the temperatures of matter and
radiation, $k_{(v',j')\rightarrow(v,j)}$ is the VT
(vibrational-translational) rate coefficient for
H+H$_2(v',j')\rightarrow$ H+H$_2(v,j)$, and $n_{(v,j)}$ describes
the distribution of rotovibrational levels,
\begin{equation}
n_{(v,j)}(T_{\rm r})=\frac{g_jn_{v}(2j+1)\exp\left(-\frac{E_{v,j}-E_{v,0}}{k_{\rm B}T_{\rm r}}\right)}{Z_v(T_{\rm r})},
\label{chi}
\end{equation}
with $g_j$ equal to $1/4$ and $3/4$ for the para- and ortho- states,
respectively, and $Z_v(T_{\rm r})$ rotational partition function for the $v^{th}$ level. It can be seen from Eqs.~(\ref{cooling})-(\ref{chi}) that
the rotational energy is distributed according to the Boltzmann law
with temperature $T_{\rm r}$ while VT coefficients depend on the matter
temperature $T_{\rm m}$.  Thus, the heat transfer function depends on two
temperatures, as a consequence of the different couplings of the
internal degrees of freedom of molecules with the radiation and the
matter: a faster coupling occuring between radiation and rotation, a
slower coupling between translation and vibration.

We explore both equilibrium and non-equilibrium cases, corresponding to
different values of $n_v$; in the former case the vibrational levels
are distributed following the Boltzmann population equation, whereas in
the latter we adopt the level populations resulting from the kinetic
model of C11. VT rate coefficients have been taken from
\cite{b5,b6}. These coefficients are given as functions of the initial
and final rotovibrational quantum numbers and of temperature allowing
the inclusion of the full sets of collisional transitions in our
calculation. The equilibrium one-temperature heat transfer using the
same rate coefficients has been compared with the analytical expression for the cooling function
given by GP98. In GP98, the gas was not embedded in any radiation field; consequently, only collisional de-excitations were considered.

\begin{figure}
\begin{center}$
\begin{array}{l}
\includegraphics[width=8.5cm]{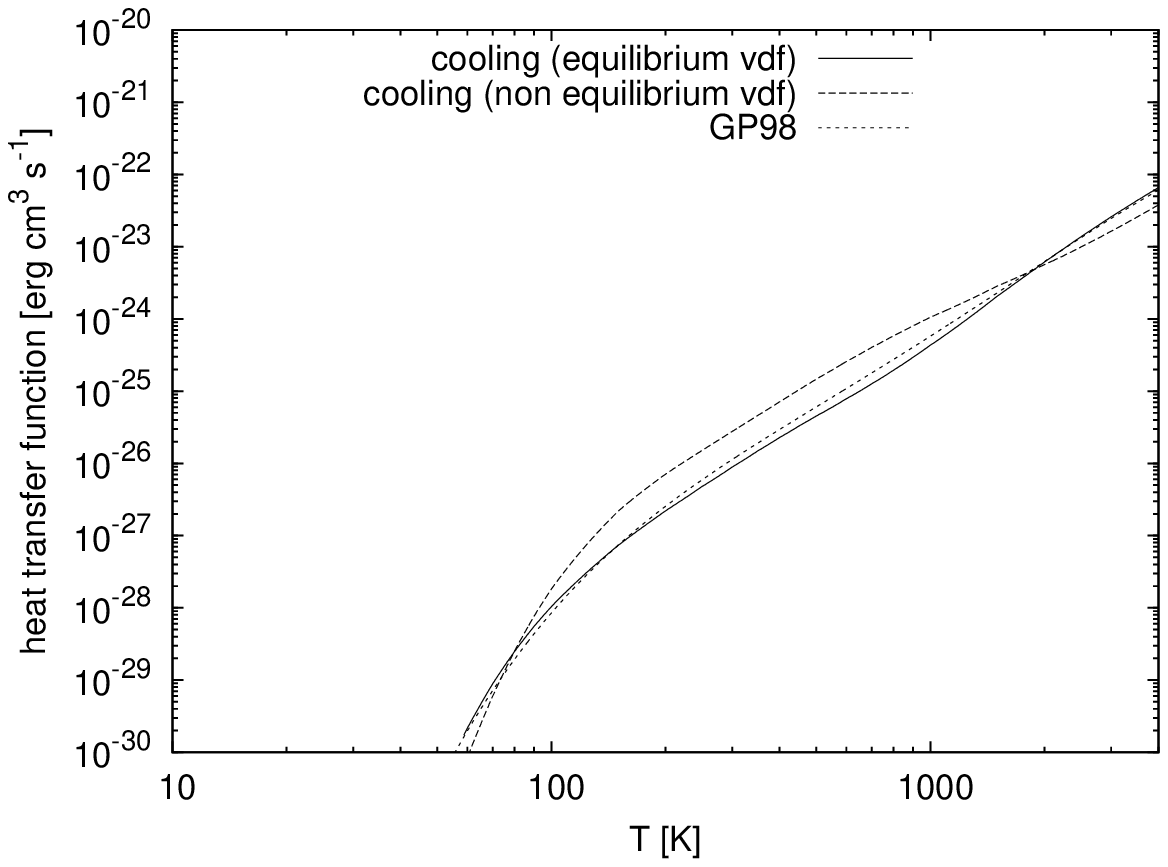} \\
\includegraphics[width=8.5cm]{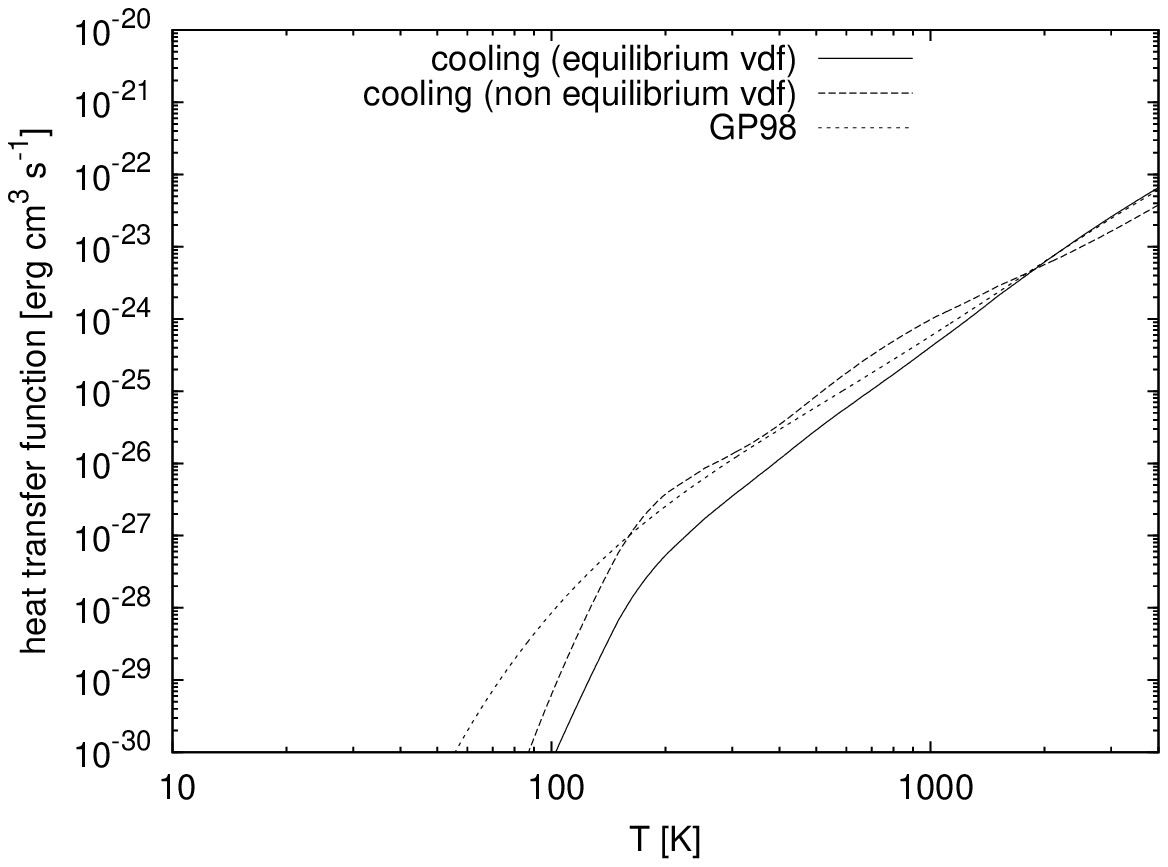} 
\end{array}$
\end{center}
\caption{Heat transfer function $(\Gamma-\Lambda)_{\rm H_2}$ as a function of the radiation temperature $T_{\rm r}$. {\it Dotted line}: GP98
fit for the cooling function; {\it solid line}: LTE calculation with present VT coefficients; {\it
dashed line}:  non-equilibrium vibrational distribution contribution.
Calculation are reported both in the one-temperature case ({\it top
panel}) and the two-temperatures one ({\it bottom panel}).}
\label{f4}
\end{figure}

As it can be seen from Figure~\ref{f4}, the non-equilibrium heat transfer function decreases more rapidly than the equilibrium one at lower temperature, because of the increased efficiency in the energy exchange due to the long sovrathermal tails in the vibrational distribution. In the two temperatures calculation, the deviation is amplified as a consequence of the strong decoupling in the energy exchange among degrees of freedom and of the different trend of the gas and radiation temperatures. The first effect, which is due to the deviation of the vibrational population from the equilibrium distribution (essentially those of the lowest levels) is seen at $z\approx 1000$, and amounts to a factor of $\sim 2$. The effect of the separation of $T_{\rm m}$ and $T_{\rm r}$ occurs at low $z$ where these two temperatures are considerably different. Consequently, in the computation of the heat transfer function at least two temperatures must be used: one is the temperature of the level population and the other is the translational temperature. This difference is evident in Eq.~(\ref{cooling}).

All heat transfer functions available in the literature are calculated
assuming a single temperature, usually set equal to $T_{\rm r}$. Such
usage cannot capture the second non-equilibrium effect described above,
since $k(T_{\rm m},T_{\rm r})$ is implicitly set equal to $k(T_{\rm
r})$. A better solution is to use in the context of early Universe
models heat transfer functions calculated including non-equilibrium effects
although fitted later as a function of a single temperature. Fits for the
non-equilibrium case using the present two-temperatures model and the
equilibrium case using the newest available data by \cite{b5,b6} are
obtained in the form:
\begin{equation}
\log_{10}~\Phi =\sum_{n=0}^N a_{n}(\log_{10} T_{\rm r})^n.
\end{equation}
The coefficients $a_n$ are listed in Table~\ref{tab2}.  For the
non-equilibrium case, the validity of the fit is up to $T\approx 100$~K
(additional data are available upon request).

\begin{table}
\begin{center}
\caption{H$_2$ heat transfer function}
\begin{tabular*}{\columnwidth}{ll}
\hline
 & Fitting coefficients \\ 
\hline
equilibrium & $a_0=-145.05$\\
            & $a_1=136.085$\\
            & $a_2=-58.6885$\\
            & $a_3=11.2688$\\
            & $a_4=-0.786142$\\
\hline
non-equilibrium & $a_0=-393.441$\\
                & $a_1=588.474$\\
                & $a_2=-380.78$\\
                & $a_3=123.858$\\
                & $a_4=-20.1349$\\
                & $a_5=1.30753$\\
\hline
\label{tab2}
\end{tabular*}
\end{center}
\end{table}

\section{Non-equilibrium reaction rates}

\begin{figure}
\begin{center}$
\begin{array}{ll}
\includegraphics[width=8.5cm]{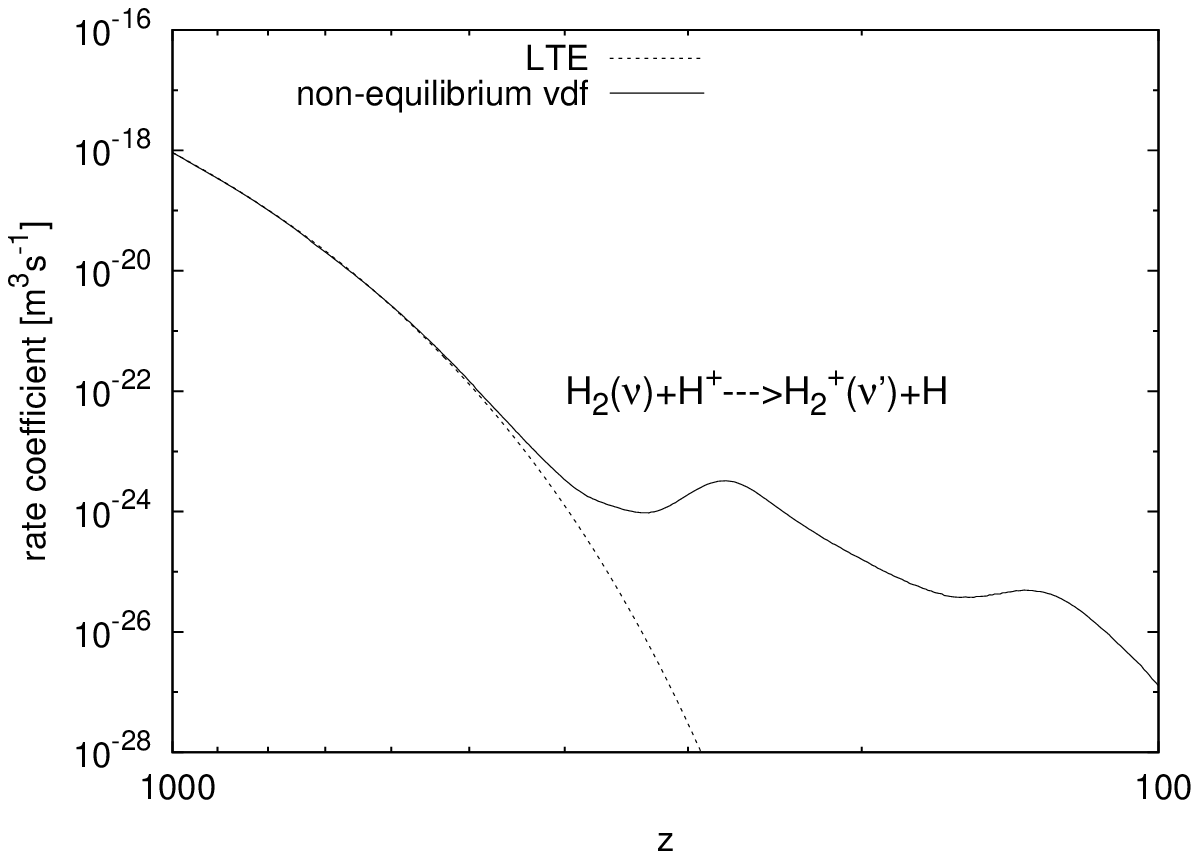} & \includegraphics[width=8.5cm]{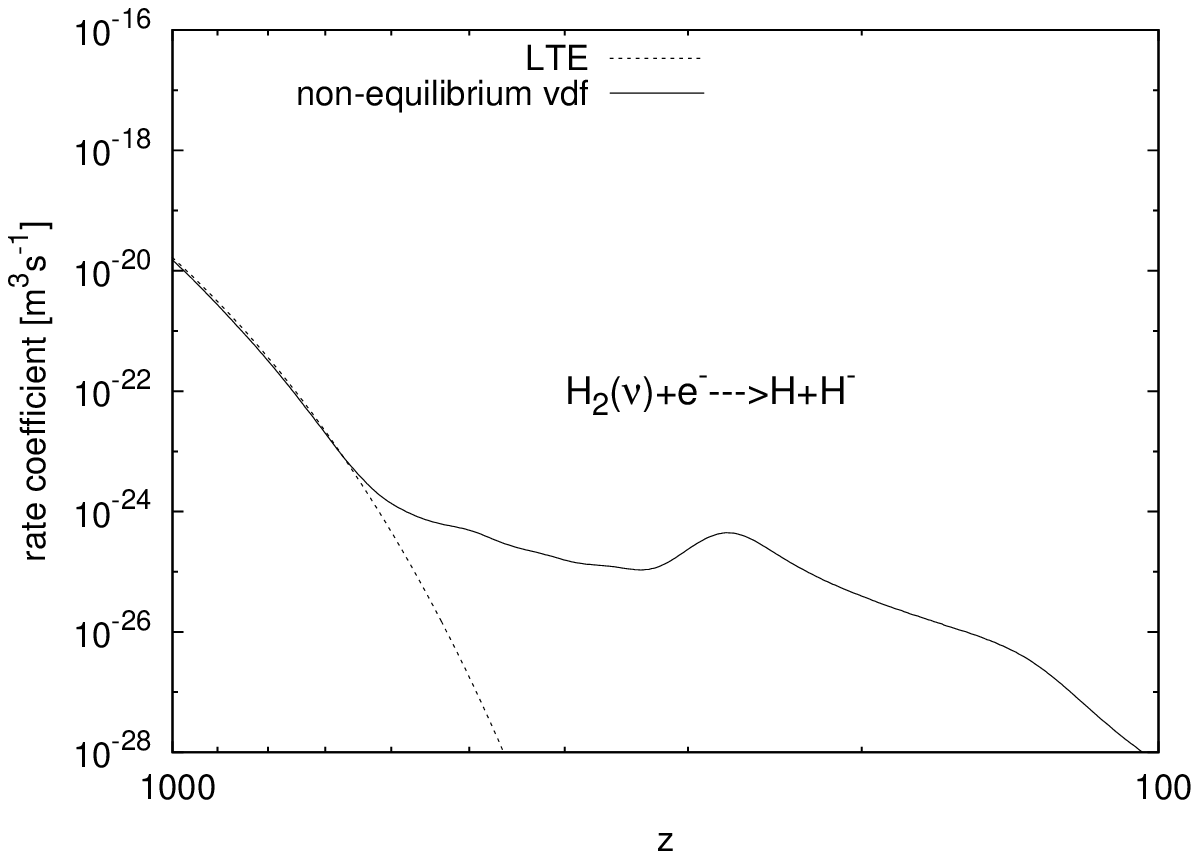} \\
\includegraphics[width=8.5cm]{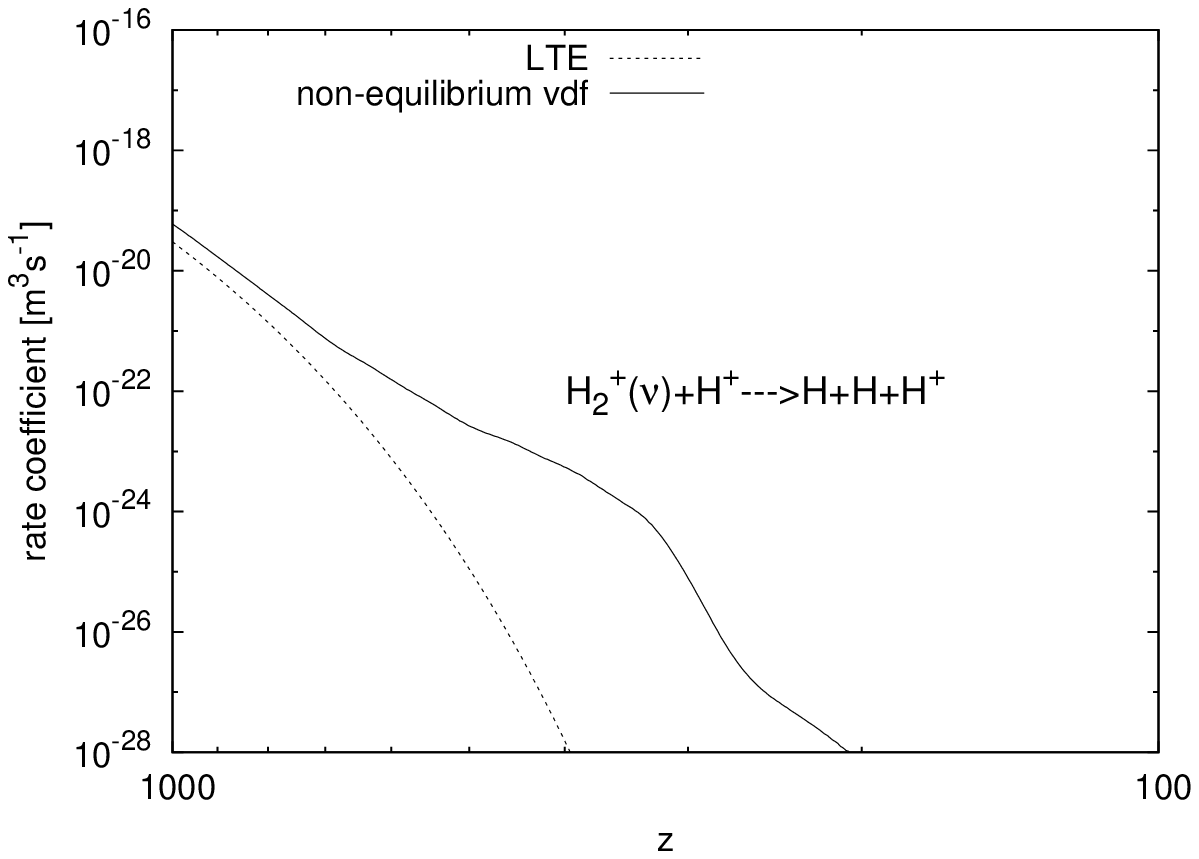} & \includegraphics[width=8.5cm]{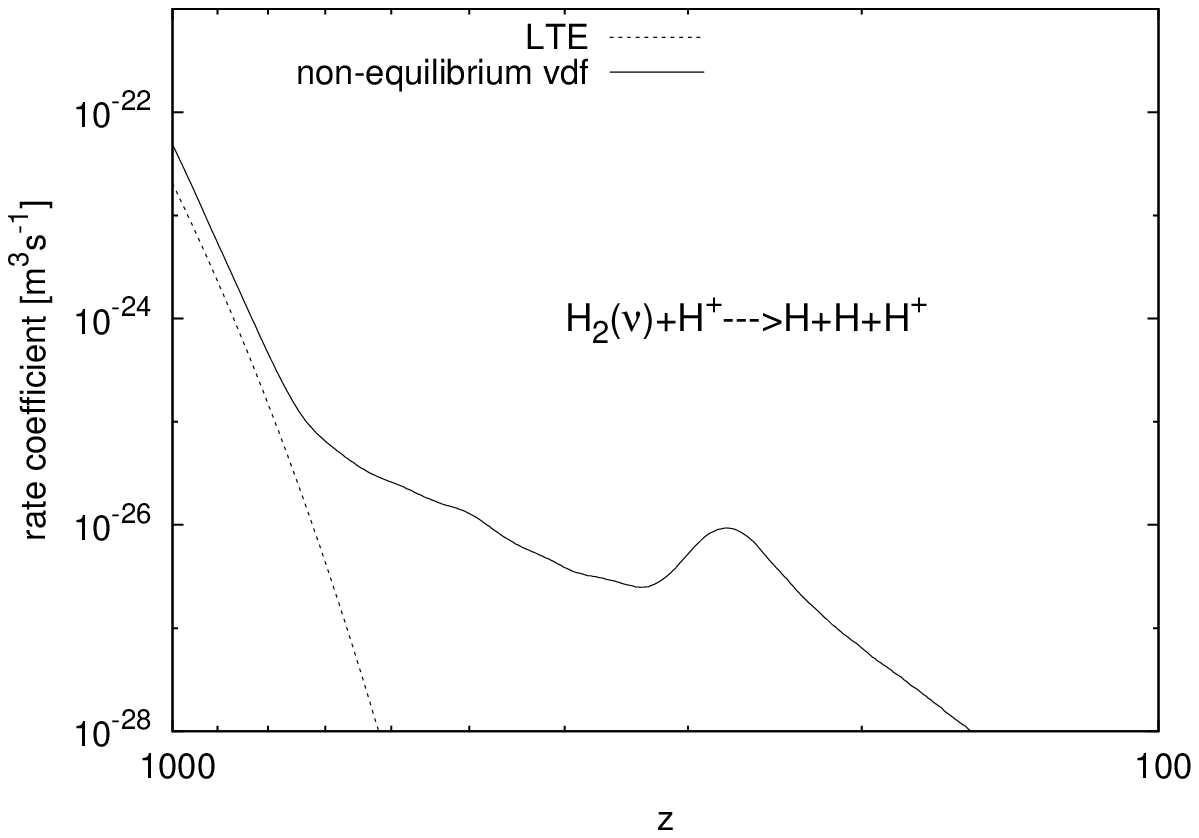} \\
\end{array}$
\end{center}
\caption{Rate coefficients as a function of $z$: LTE approximation
({\it dashed line}, fit by Coppola et al.~2011a) and non-equilibrium
vibrational distribution function ({\it solid line}).}
\label{f5}
\end{figure}

Using the real non-equilibrium vibrational distributions, vibrationally
resolved rate coefficients have been recomputed and compared with the
corresponding LTE fits by C11. In Figure~\ref{f5} the results for the following
processes introduced in the model are shown, both for H$_2$ and
H$_2^+$: (1) H$_2(v)$/H$^+$ charge transfer (2) H$_2(v)$/e$^-$
dissociative attachment (3) H$_2^+(v)$ dissociation by collisions with H (4) H$_2(v)$ dissociation by
collisions with H$^+$.  The LTE fit for dissociative attachment of
H$_2$ is taken from Capitelli et al.~(2007). Strong deviations from the LTE fits can be noted, due to the
non-equilibrium pattern, especially at low $z$, where the hypothesis of
Boltzmann distribution of the vibrational level manifold fails. The
peak at $z\approx 300$ corresponds to that on H$_2^+$ (and consequently
to H$_2$, via the process of charge transfer with H$^+$). It should be
noted that, in the case of H$_2$ dissociative attachment, the
non-equilibrium calculation follows the trend reported by Capitelli et
al.~(2007), where a simplified model for the non-equilibrium
distribution was  assumed.

\clearpage
\section{Spectral distortions of the CBR}

The photons created in the H$_2$ formation process produce a
distortion of the black-body spectrum of the cosmic background
radiation (CBR). Since the maximum production of H$_2$ occurs in
vibrationally excited states by associative detachment in collisions of
H and H$^-$ at redshifts below $z\approx 100$ (see Coppola et
al.~(2011a)), the emission of rovibrational transitions with wavelength
$\lambda \approx 2$~$\mu$m is redshifted today at $\lambda\approx
100$--200~$\mu$m, in the Wien part of the CBR.  An early estimate of
this distortion, based on the rovibrational-resolved associative
detachment cross sections   computed by Bieniek \& Dalgarno~(1979), was
made by Khersonskii~(1982) while Shchekinov \& \'Ent\'el~(1984)
developed a model for the molecular hydrogen distortion due to
secondary heating processes. We reconsider here the process of
vibrational emission of primordial H$_2$ molecules with our updated
chemical network and with a fully kinetic treatment of the level
populations of H$_2$.

For each transition from an upper level $v_u$ to a lower level $v_l$,
with level populations $n_u$ and $n_l$ and degeneracy coefficients
$g_u$ and $g_l$, the relative perturbation in the CBR at the present
time is given by
\begin{equation}
\left.\frac{\Delta J_\nu}{J_\nu}\right|_{z=0}=
[S(z_{\rm int})-1]\tau(z_{\rm int}),
\end{equation}
where
\begin{equation}
S(z_{\rm int})=\left[\frac{g_u n_l(z_{\rm int})}{g_l n_u(z_{\rm int})}-1\right]^{-1}
\left\{\exp\left[\frac{h\nu_{ul}}{kT_{\rm r}(z_{\rm int})}\right]-1\right\}
\label{source}
\end{equation}
is the source function, and 
\begin{equation}
\tau(z_{\rm int}) = \frac{c^3}{8\pi\nu_{ul}^3}A_{ul}\frac{g_u}{g_l}
\left[1-\frac{g_ln_u(z_{\rm int})}{g_un_l(z_{\rm int})}\right]
\frac{n_l(z_{\rm int})}{H_z(z_{\rm int})},
\label{tau}
\end{equation}
is the redshift-integrated optical depth (see e.g. Appendix A of
Bougleux \& Galli~1997). In Eqs.~(\ref{source}) and (\ref{tau}), $z_{\rm int}$
is the interaction redshift, at which the observed frequency $\nu$ is equal to 
the redshifted frequency $\nu_{ul}$ of the transition, i.e.
\begin{equation}
\nu(1+z_{\rm int})=\nu_{ul},
\end{equation}
$A_{ul}$ are the Einstein coefficients, and $H_z$ is the Hubble function
\begin{equation}
H_z=H_0[\Omega_r(1+z)^4+\Omega_m(1+z)^3+\Omega_k(1+z)^2+\Omega_\Lambda]^{1/2} 
\end{equation}
(see Coppola et al.~(2011a) for a definition of the cosmological constants and 
their adopted values).

\begin{figure}
\includegraphics[width=8cm]{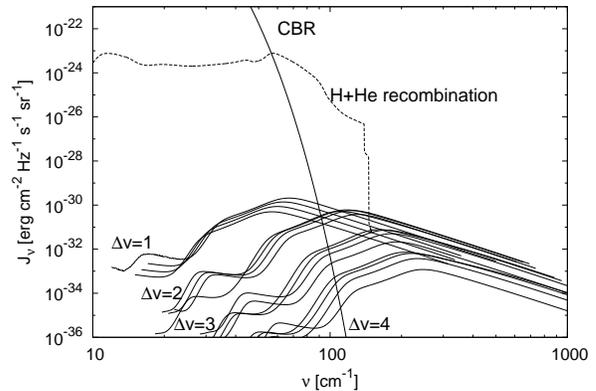} 
\caption{CBR spectrum at $z$=0 together with spectral distortions due
to H and He recombinations (\cite{b255}, \cite{b256}) and to non-equilibrium vibrational molecular
transitions for H$_2$. The contribution of multiquantum transitions up
to $\Delta v=4$ are shown.}
\label{f6}
\end{figure}

Figure~\ref{f6} shows the emission produced by H$_2$ transitions with
$\Delta v=1$, $2$, $3$ and $4$ in the frequency range
$\nu=10$--1000~cm$^{-1}$, corresponding to wavelengths
$\lambda=10$~$\mu$m--1~mm. To avoid confusion, only the first 4
transitions for each $\Delta v$ are shown (i.e., $v=1\rightarrow 0$,
$2\rightarrow 1$, $3\rightarrow 2$ and $4\rightarrow 3$ for $\Delta
v=1$, etc.). The figure also shows the CBR in the Wien region, and, for
comparison, the spectral features produced by the cosmological
recombination of H and He (e.g. Chluba \& Sunyaev
2007,2008, \cite{b255}, \cite{b256}).  The latter are mainly formed by redshifted
Ly-$\alpha$ and two-photon transitions of H and the corresponding lines
from He (Chluba \& Sunyaev~2010 have recently produced new results for this last contribution).

The difficulty of detecting spectral distortions in the Wien side of
the CBR, in the presence of an infrared background (both Galactic and
extragalactic) several orders of magnitude brighter, have been
discussed by Wong et al.~(2006). While a direct detection appears
challenging (see also Schleicher et al.~2008), we stress that an excess
of photons over the CBR at wavelengths shorter than the peak could
represent a significant contribution to several photodestruction
processes, as shown by Switzer \& Hirata~(2005) for the
photoionization of Li and Hirata \& Padmanabhan~(2006) for the
photodetachment of H$^-$. Another possibility is fluorescence, i.e.
the absorption of the short-wavelengths non-thermal photons by atoms or
molecules followed by re-emission at longer wavelenghts in the
Rayleigh-Jeans region of the CBR, as suggested by Dubrovich \&
Lipovka~(1995). These issues will be addressed elsewhere.

\section{Conclusions}

We have considered several vibrational non-equilibrium effects on the
chemistry and physics of early Universe.  Although our present
calculations raise several questions about the very use of the concept
of temperature in the early Universe, the most important issue
concerning thermal transfer and chemical reactivity has been
addressed.  Our calculations show that the differences between the
excitation, translation and radiation temperatures can affect the
heat transfer functions of important species, an effect here demonstrated for
H$_2$.  Our results underline the necessity to fully include the
consequences of the temperature separations which occur at different
epochs in the chemical and physical evolution of the early Universe.

Excitation temperatures appear to be higher than radiation temperature
at low $z$; this ``chemical'' pre-heating should be considered while
modeling the formation of galaxies, together with virialization heating
and other physical mechanisms usually suggested (e.g. \cite{b16},
\cite{b17}). We have also assessed the hypothesis of steady-state for
the vibrational distribution presenting a reduced kinetic model, and
calculated the deviations between vibrational temperatures of ortho-
and para- states as a first step towards a full non-equilibrium
rotovibrational kinetics. Heat transfer functions are calculated for both equilibrium and non-equilibrium cases, considering also a novel two-temperatures approach that takes into account the different rates of energy exchange among molecular degrees of freedom. For pure vibrational transitions, the critical density is greater than the baryon density, so that the hypothesis of non-equilibrium is also valid  at higher $z$.
Resolving rotations and vibrations gives different results, making the limit for $z$ lower (e.g. for the $(0,2)\rightarrow (0,0)$ transition the critical density is about $\approx 2.7$x$10^7~\mathrm m^{-3}$ at $z \approx 500$.)

We have evaluated the effects of vibrational non-equilibrium on 
reaction rates. A general increase has been
pointed out because of the formation of long sovrathermal tail in the
vibrational distribution, especially at low $z$; this evidence should
be added to the increase of rate coefficients due to the inclusion
of the entire vibrational manifold, that by itself can affect in a deep
way the fate of the system modeled (as described by \cite{b20} for the
dissociative attachment process).  

We have computed the spectral deviations to the CBR due to the
non-equilibrium level populations, considering all the transitions.
Although the present Planck experiment and the upcoming James Webb
Space Telescope (JWST) are able to detect galaxies at high redshift, a
direct observation of this effect is challenging; for this reason, an
alternative study of the non-thermal vibrational photons on the
photochemical pathways of atomic and molecular kinetic should be
undertaken.

\section*{Acknowledgments}
We are grateful to Jens Chluba for having made available his data and for helpful discussions.
CMC and SL acknowledge financial support of MIUR-Universit\`a degli
Studi di Bari, (\textquotedblleft fondi di Ateneo 2011
\textquotedblright). This work has also been partially supported by the
FP7 project ''Phys4Entry'' - grant agreement n. 242311. JT acknowledges
support from ERC Advanced Investigator Project 267219.

\label{lastpage}

\end{document}